\documentstyle[aps,floats,epsf,twocolumn]{revtex}

\begin{document} \draft

\title{Vortex state in a doped Mott insulator}

\author{M. Franz and Z. Te\v{s}anovi\'c}
\address{Department of Physics and Astronomy, Johns Hopkins University,
Baltimore, MD 21218
\\ {\rm(\today)}
}
%\maketitle
%
%\begin{abstract}
\address{~
\parbox{14cm}{\rm
\medskip
We analyze the recent vortex core spectroscopy data on cuprate
superconductors and discuss what can be learned from them about the nature 
of the ground state in these compounds. We argue that the data are
inconsistent with the assumption of a simple metallic 
ground state and exhibit characteristics of a doped Mott insulator. 
A theory of the vortex core in such a doped Mott insulator is 
developed based on the U(1) gauge field slave boson model. In the limit of
vanishing gauge field stiffness such theory predicts two types of singly 
quantized vortices: an insulating ``holon'' vortex in the underdoped 
and metallic ``spinon'' vortex in the overdoped region of the phase diagram.
We argue that the holon vortex exhibits a pseudogap excitation spectrum in 
its core qualitatively consistent with the existing experimental data on 
Bi$_2$Sr$_2$CaCu$_2$O$_8$. As a test of this theory we propose that spinon 
vortex with metallic core might be observed in the heavily overdoped samples. 
}}
%\end{abstract}
\maketitle

%\pacs{74.60.-w,74.60.Ec,74.72.-h}

%
\narrowtext

\section{Introduction}
Nature of the ground state as a function of doping remains one of the recurring
unresolved issues 
in the theory of high-$T_c$ cuprate superconductors. The problem is
partly due to formidable difficulties related to the theoretical description of
doped Mott insulators and partly due to experimental hurdles in accessing
the normal state properties in the $T\to 0$ limit because of the intervening
superconducting order. Probes that suppress superconductivity and 
reveal the properties of the underlying ground state are therefore of 
considerable value. So far only pulsed
magnetic fields\cite{ando1} in excess of $H_{c2}$ and impurity doping
beyond the critical concentration\cite{lemberger1} have been used
towards this goal. Here we argue that the vortex core
spectroscopy performed using scanning tunneling microscope (STM) 
can provide new insights into the nature of the ground state in cuprates.
We analyze the existing experimental data\cite{maggio1,renner1,pan1,pan2} 
and conclude that they imply strongly correlated ``normal'' ground state,
presumably derivable from a doped Mott insulator. We then 
develop a theoretical framework for the problem of 
tunneling in the vortex state of such a doped Mott insulator.

In the vortex core the superconducting order parameter is locally suppressed 
to zero and the region within a coherence length $\xi$ from its center can
be to the first approximation thought of as normal. Spectroscopy of the
vortex core therefore provides information on the normal state electronic
excitation spectrum in the $T\to 0$ limit. More accurately, the core 
spectroscopy reflects the spectrum in the spatially non-uniform situation
where the order parameter amplitude rapidly varies in response to the
singularity in the phase imposed by the external magnetic field. In order to
extract useful information regarding the underlying ground state
from such measurements a detailed understanding of the vortex core 
physics is necessary. So far the problem has been addressed using the weak 
coupling approach based on the 
Bogoliubov-de Gennes theory generalized to the $d$-wave symmetry of 
the order parameter\cite{soininen1,wang1,franz1,kita1}, and semiclassical
calculations\cite{volovik1,maki1,ichioka1}.
The early theoretical debate focused on the existence
or absence of  the vortex core bound states 
\cite{maki1,franz2,himeda1}. This debate, now resolved in favor
of absence of any bound states in pure $d_{x^-y^2}$ state
\cite{franz1,kita1,resende1}, has somewhat eclipsed the possibly
more important issues related to the nature of the ground state 
in cuprates.

The body of work based on mean field, weak coupling calculations
\cite{wang1,franz1,kita1,ichioka1} yields results for the
local density of states in the vortex core which exhibit two generic features:
(i) the coherence peaks (occurring at $E=\pm\Delta_0$ in the bulk) are 
suppressed,
with the spectral weight transferred to a (ii) broad featureless peak 
centered around the zero energy. 
%Both features are easily understood on 
%physical grounds. The smearing of the coherence peaks results from the gap
%not being sharp on the length-scale $\xi$ in the core and the broad peak is a
%remnant the bound states that would exist in the $s$-wave 
%vortex\cite{caroli1} strongly hybridized into the continuum states 
%present in a $d$-wave superconductor. 
Here we wish to emphasize the heretofore little appreciated fact that 
these features are {\em qualitatively inconsistent}
with the existing experimental data on cuprate superconductors. STM
spectroscopy on Bi$_2$Sr$_2$CaCu$_2$O$_8$
(BSCCO) at 4.2K indicates a ``pseudogap'' spectrum in the vortex core with the 
spectral weight from the coherence peaks at $\pm\Delta_0\simeq 40$meV
transferred to {\em high energies}, and no peak whatsoever
around $E=0$\cite{renner1}. Recent high resolution 
data on the same compound\cite{pan2}
confirmed these findings down to 200mK and found evidence for weak bound
states at $\pm7$meV. Experiments on YBa$_2$Cu$_3$O$_7$ (YBCO)\cite{maggio1}
also indicate low energy bound states, but are somewhat more difficult to
interpret because of the high zero-bias conductance of unknown origin
appearing even in the absence of magnetic field. 

The fundamental discrepancy between the theoretical predictions and 
the experimental findings strongly suggests that models based on a simple
weak coupling theory break down in the vortex core. The pseudogap observed 
in the core hints that the underlying ground state revealed
by local suppression of the superconducting order parameter is 
a doped Mott insulator and not a conventional metal. Taking into account 
the effects of strong correlations appears to be necessary to
consistently describe the physics of the vortex core. 
Conversely, studying the vortex core physics could provide information
essential for understanding the nature of the underlying ground state 
in cuprates.  

The first step in this direction was taken by Arovas {\em et al.}
\cite{arovas1} who proposed that within the framework of the SO(5) theory
\cite{zhang1} vortex cores could become antiferromagnetic (AF). 
They found that such AF cores can be stabilized
at low $T$ but only in the close vicinity of the bulk AF phase. In contrast,
experimentally the pseudogap in the core is found to persist 
into the overdoped 
region\cite{renner1}. More recently microscopic calculations within the same 
model\cite{andersen1} revealed electronic excitations in such AF cores with 
behavior roughly resembling the experimental data. Quantitatively, 
however, these spectra exhibit asymmetric
shifts in the coherence peaks (related to the fact that spin gap in the AF 
core is no longer tied to the Fermi level) not observed experimentally. These
discrepancies suggest that generically cores will not exhibit the true AF 
order. Finally, these previous approaches are still of the 
Hartree-Fock-Bogoliubov type and cannot be expected to properly capture the
effects of strong correlations.

\begin{figure}[t]
\epsfxsize=8.5cm
\epsffile{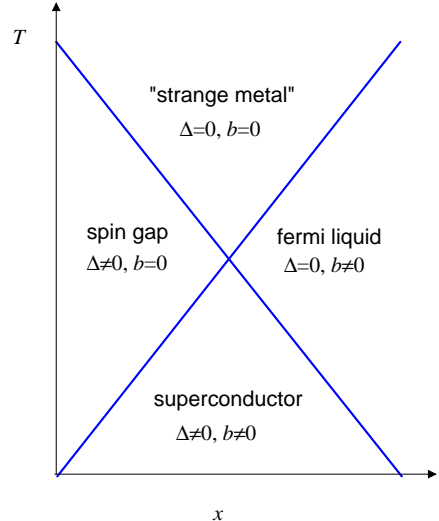}
\caption[]{Schematic phase diagram of the system with spin-charge
separation in the doping$-$temperature plane, as applied to cuprate
superconductors. }
\label{fig1}
\end{figure}

Here we consider a model for the vortex core based on a version of 
the U(1) gauge field slave boson theory formulated recently by 
Lee\cite{dhlee1}. Originally proposed by Anderson\cite{anderson1} 
the slave boson theory was formulated 
to describe strongly correlated electrons in the CuO$_2$ planes of the 
high-$T_c$ cuprates. Various versions of
this theory have been extensively discussed in the 
literature \cite{baskaran1,ruckenstein1,kotliar1,affleck1,lee1}. 
Interest in spin-charge  separated 
systems revived recently\cite{wen1,lee2,balents1,dhlee1} due to the
realization that it provides a natural description of the pseudogap phenomenon
observed in the underdoped cuprates. The common ingredient in these theories
is ``splintering'' of the electron into quasiparticles carrying its spin
and charge degrees of freedom. Within the theories based on Hubbard and 
$t$-$J$ models this splintering is formally implemented by the
decomposition of the electron creation operator
\begin{equation}
c^\dagger_{i\sigma}=f^\dagger_{i\sigma} b_i
\label{}
\end{equation}
into a fermionic spinon $f_{i\sigma}$ and bosonic holon $b_i$. The local
constraint of the single occupancy $b^\dagger_ib_i+
f^\dagger_{i\sigma}f_{i\sigma}=1$ is enforced by a fluctuating U(1) gauge 
field ${\bf a}$. The mean field phase diagram is known to contain
four phases distinguished 
by the formation of spinon pairs, $\Delta_{ij}=\langle\epsilon_{\sigma\sigma'}
f^\dagger_{i\sigma}f^\dagger_{j\sigma'}\rangle$, and Bose-Einstein condensation
of the individual holons $b=\langle b_i\rangle$\cite{lee1}, and is illustrated
 in Figure (\ref{fig1}).

The effects of magnetic field on such spin-charge separated system is
most conveniently studied in the framework of an effective Ginzburg-Landau 
(GL) theory for the condensate fields $\Delta$ and $b$. The corresponding 
effective action can be constructed\cite{sachdev1,lee3}
based on the requirements of local gauge invariance with respect to 
the physical electromagnetic vector potential ${\bf A}$ and the internal
gauge field ${\bf a}$:
\begin{eqnarray}
f_{\rm GL}&=&|(\nabla-2i{\bf a})\Delta|^2 +{r_\Delta}|\Delta|^2 +
{1\over 2}{u_\Delta}|\Delta|^4\nonumber \\
&+&|(\nabla-i{\bf a}-ie{\bf A})b|^2 +r_b|b|^2 +{1\over 2} u_b|b|^4 
 +v|\Delta|^2|b|^2 \nonumber \\
&+& {1\over 8\pi}(\nabla\times{\bf A})^2 +f_{\rm gauge}.
\label{fgl1}
\end{eqnarray}
The factor of 2 in the spinon gradient term reflects the fact that {\em pairs}
of spinons were assumed to condense. $f_{\rm gauge}$ describes the dynamics
of the internal gauge field ${\bf a}$. We note that unlike the physical 
electromagnetic field ${\bf A}$ the gauge field ${\bf a}$ has no independent dynamics
in the underlying microscopic model since it serves only to enforce a 
constraint. Sachdev\cite{sachdev1} and Nagaosa and Lee\cite{lee3} assumed
that upon integrating out the microscopic degrees of freedom a term
\begin{equation}
f_{\rm gauge}={\sigma\over2}(\nabla\times{\bf a})^2
\label{fgauge}
\end{equation}
is generated in the free energy. They then analyzed vortex solutions of
the free energy (\ref{fgl1}) and came to the conclusion that two types of
vortices are permissible: a ``holon vortex'' with the singularity in the 
$b$ field and a ``spinon vortex'' with the singularity in the
 $\Delta$ field. Because holons
carry electric charge $e$ the holon vortex is threaded by electronic flux
quantum $hc/e$, i.e. twice the conventional 
superconducting flux quantum $\Phi_0=hc/2e$.
Spinons on the other hand condense in pairs, and the spinon vortex therefore
carries flux $\Phi_0$.
Stability analysis then implies that spinon vortex will be stable over the
most of the superconducting phase diagram, while the 
$hc/e$ holon vortex can be stabilized
only in the close vicinity of the phase boundary on the underdoped side
\cite{sachdev1,lee3}. 
This is a direct consequence of the fact that singly quantized vortices are 
always energetically favorable\cite{abrikosov1,fetter1}.

As far as the electronic excitations are concerned, the spinon vortex is
virtually indistinguishable from the vortex in a conventional weak coupling
mean field
theory: the spin gap $\Delta$, which gives rise to the gap in the electron
spectrum, vanishes in the core. Consequently, the vortex state based on the
results of Sachdev-Nagaosa-Lee (SNL) theory\cite{sachdev1,lee3} does not
exhibit the pseudogap in the core and suffers 
from the same discrepancy with the experimental data as the weak coupling
theories\cite{soininen1,wang1,franz1,kita1} based on the conventional Fermi 
liquid description. Moreover, no evidence exists at present
for stable doubly quantized holon vortices predicted by SNL. What is
needed to account for the experimental data is a {\em singly quantized
holon vortex} stable over the large portion of the superconducting phase in the
phase diagram of Figure \ref{fig1}. In the core of such a holon vortex the spin
gap $\Delta$ remains finite and leads naturally
to the pseudogap excitation spectrum. In what follows we show that under
certain conditions the free energy (\ref{fgl1}) permits precisely such 
solution. 

The results of the SNL theory are predicated upon the assumption that the 
``stiffness'' $\sigma$ of the gauge field is relatively large and that singular
configurations in which $\nabla\times{\bf a}$ contains a full flux quantum
through an elementary plaquette are prohibited. Consider now a precisely 
opposite physical situation, allowing unconstrained fluctuations in 
${\bf a}$. This amounts
to the assumption that  the $f_{\rm gauge}$ term (\ref{fgauge})
can be neglected in (\ref{fgl1}), i.e. $\sigma\to 0$. Physically this
corresponds to the ``extreme type-I'' limit of the GL ``superconductor''
(\ref{fgl1}) with respect to fluctuations in ${\bf a}$. Based on Elitzur's 
theorem\cite{elitzur1}  Nayak\cite{nayak1} recently argued that
the exact local U(1) symmetry of the model cannot be broken, implying absence
of the phase stiffness term (\ref{fgauge}) at all energy scales. Our assumption
therefore appears reasonable and in Section III. we shall give a more thorough
discussion of the significance of the $f_{\rm gauge}$ term for the vortex
solutions of interest here. For the time being we shall assume that 
$f_{\rm gauge}$ can be neglected and explore physical consequences of the
resulting theory. 

$f_{\rm GL}$ given by Eq.\ (\ref{fgl1}) is quadratic in ${\bf a}$ and with 
the $\nabla\times{\bf a}$ term absent the gauge
fluctuations can be trivially integrated out. Within the closely related 
microscopic model this procedure has been recently implemented by 
Lee\cite{dhlee1}. The resulting effective free energy density reads
\begin{eqnarray}
f&=& f_{\rm amp} +{\rho_\Delta^2\rho_b^2\over 4\rho_\Delta^2+\rho_b^2}
(\nabla\phi-2\nabla\theta+2e{\bf A})^2 \nonumber \\
&+&{1\over 8\pi}(\nabla\times{\bf A})^2,
\label{feff}
\end{eqnarray}
where we have set $\Delta=\rho_\Delta e^{i\phi}$, $b=\rho_b e^{i\theta}$, and 
\begin{eqnarray}
f_{\rm amp}&=&(\nabla\rho_\Delta)^2 +{r_\Delta}\rho_\Delta^2 +
{1\over 2}{u_\Delta}\rho_\Delta^4\nonumber \\
&+&(\nabla\rho_b)^2 +r_b\rho_b^2 +{1\over 2} u_b\rho_b^4 
 +v\rho_\Delta^2\rho_b^2
\label{famp}
\end{eqnarray}
is the amplitude piece. The most important feature of the effective free
energy (\ref{feff}) is that it no longer depends on the individual phases
$\phi$ and $\theta$ but only on their particular combination
\begin{equation}
\Omega=\phi-2\theta.
\label{omega}
\end{equation}
Since the physical superconducting order parameter $\Psi=\Delta^*b^2=
\rho_\Delta\rho_b^2 e^{-i(\phi-2\theta)}$ it is reasonable 
to identify $\Omega$ with the phase of a {\em Cooper pair}. Physically, the
unconstrained fluctuations of the gauge field in Eq.\ (\ref{fgl1}) resulted in
partial restoration of the original electronic degrees of freedom in Eq.\ 
(\ref{feff}). In the underlying microscopic model this means that on long
length scales spinons and holons are always confined, in agreement with
Elitzur's theorem \cite{elitzur1,nayak1}. On lengthscales shorter than
the confinement length, such as inside the vortex core, spinons and holons
can still appear locally decoupled. In the present effective theory 
this aspect is reflected by two amplitude degrees of freedom
present in (\ref{feff}). More detailed discussion of these issues is given in 
Refs.\ \cite{dhlee1,nayak1}.  

We have thus arrived at an effective theory of a spin-charge separated
system containing one phase degree
of freedom $\Omega$ and two amplitudes, $\rho_\Delta$ and $\rho_b$. Deep in the
superconducting phase,
where both amplitudes are finite, the physics of (\ref{feff}) will be very
similar to that of a conventional GL theory. In the situations where
the superconducting order parameter $\Psi$ is strongly suppressed, such as
in the vortex core, near an impurity or a wall, the new theory has an extra
degree of richness, associated with the fact that it is sufficient (and
generally preferred by the energetics) when only {\em one}
 of the two amplitudes 
is suppressed. Since the two amplitudes play very different roles in the 
electronic excitation spectrum, the effective theory (\ref{feff})
will lead to a number of nontrivial effects.

To illustrate this consider what will happen in the core of a superconducting
vortex. Under the
influence of the magnetic field the phase $\Omega$ will develop a singularity
such that $\nabla\Omega \sim 1/r$ close to the vortex center. For 
the free energy to remain finite the amplitude prefactor in the second
term of Eq.\ (\ref{feff}) must vanish for $r\to 0$. 
This is analogous to $|\Psi|$ vanishing
in the core of a conventional vortex. In the present case, 
however, it is sufficient when the product $\rho_\Delta\rho_b$ vanishes. Since 
suppressing any of the two amplitudes costs condensation energy, in general
only one amplitude will be driven to zero. Which of the two is suppressed
will be determined by the energetics of the amplitude term (\ref{famp}).
On general grounds we expect that the state in the vortex core will be
the same as the corresponding bulk ``normal'' state obtained by raising 
temperature above $T_c$. Thus, very crudely, we expect that holon vortex 
will be stable in the underdoped while the spinon vortex will be stable in the
overdoped region of the phase diagram Figure \ref{fig1}. 

An important point by which our approach differs from the SNL theory
is that in the present theory {\em both} types of vortices carry the 
{\em same} superconducting flux quantum $\Phi_0$ and thus compete
on equal footing. This is a direct consequence of our assumption of
the vanishing phase stiffness $\sigma$. 

In what follows we study in detail the vortex solutions of the free energy
(\ref{feff}). Our main objective is to obtain the  precise estimates for 
the energy of the two types of vortices as a function of temperature and 
doping and deduce the corresponding phase diagram for the state inside
the vortex core. We show that for generic parameters in (\ref{feff})
the singly quantized holon vortex with a pseudogap spectrum in the core can 
be stabilized over a large portion of the superconducting phase, as
required by the experimental constraints discussed above.

\section{Solution for a single vortex}

\subsection{General considerations}

In order to provide a more quantitative discussion we now adopt some 
assumptions
about the coefficients entering the free energy (\ref{feff}). We assume that 
\begin{equation}
r_i=\alpha_i(T-T_i), \ \ i=b,\Delta,
\label{ri}
\end{equation}
where $T_i$ are corresponding ``bare'' critical temperatures, which we
assume depend on doping concentration $x$ in the following way:
\begin{equation}
T_\Delta = T_0(2x_m-x), \ \ \
T_b      = T_0x.
\label{tc}
\end{equation}
Here $x_m$ denotes the optimal doping and $T_0$ sets the overall temperature
scale. We furthermore assume that $u_i$ and $v$ are all positive and
independent of doping and temperature. It is easy to see that such choice 
of parameters qualitatively reproduces the bulk phase diagram of cuprates
in the $x$-$T$ plane shown in  Figure \ref{fig1}. 
The effect of the $v$-term is to suppress $T_c$ from 
its bare value away from the optimal doping. In real systems fluctuations
will lead to additional suppression of $T_c$ which we do not consider here. 

In the absence of perturbations the bulk values of the amplitudes are 
given by 
\begin{eqnarray}
\bar{\rho}_\Delta^2 &=& -({r_\Delta} u_b-r_bv)/D, \nonumber \\
\bar{\rho}_b^2 &=& -(r_b{u_\Delta} -{r_\Delta} v)/D,
\label{rho}
\end{eqnarray}
with $D=u_b{u_\Delta}-v^2$. 
In analogy with conventional GL theories we may define coherence
lengths for the two amplitudes\cite{sachdev1}
\begin{eqnarray}
\xi_\Delta^{-2} &=& -({r_\Delta}-r_bv/u_b), \nonumber \\
\xi_b^{-2}     &=& -(r_b-{r_\Delta} v/{u_\Delta}), 
\label{xi}
\end{eqnarray}
one of which always diverges at $T_c$ as $(T-T_c)^{-1/2}$. 

Minimization of the free energy (\ref{feff}) with respect to the vector 
potential ${\bf A}$ yields an equation 
\begin{equation}
\nabla\times\nabla\times{\bf A}=e\rho_s(\nabla\Omega-2e{\bf A}),
\label{lona}
\end{equation}
where 
\begin{equation}
\rho_s={4\rho_\Delta^2\rho_b^2\over 4\rho_\Delta^2+\rho_b^2}
\label{rhos}
\end{equation}
is the effective superfluid density. The term in brackets can be 
identified as twice the conventional superfluid velocity 
$${\bf v}_s={1\over 2}\nabla\Omega-e{\bf A}.$$ 
Making use of the Ampere's law
$4\pi{\bf j}=\nabla\times{\bf B}$ we see that Eq.\ (\ref{lona}) specifies
the supercurrent in terms superfluid density and velocity:
${\bf j}=2e\rho_s{\bf v}_s$. Minimization of (\ref{feff}) with respect
to $\Omega$ then implies $\nabla\cdot{\bf j}=0$; the supercurrent is 
conserved.

Minimizing the free energy (\ref{feff}) with respect to the amplitudes results
in the pair of coupled GL equations:
\begin{mathletters}
\label{gl:all}
\begin{equation}
-\nabla^2\rho_\Delta + {r_\Delta}\rho_\Delta + {u_\Delta}\rho_\Delta^3 + v\rho_b^2\rho_\Delta
+{4\rho_\Delta^2\rho_b^2\over (4\rho_\Delta^2+\rho_b^2)^2}{\bf v}_s^2 = 0, \label{gl:a}
\end{equation}
\begin{equation}
-\nabla^2\rho_b + r_b\rho_b + u_b\rho_b^3 + v\rho_\Delta^2\rho_b
+{16\rho_\Delta^2\rho_b^2\over (4\rho_\Delta^2+\rho_b^2)^2}{\bf v}_s^2 = 0. \label{gl:b}
\end{equation}
\end{mathletters}
We are interested in the behavior of the amplitudes in the vicinity of the 
vortex center. In this region, for a strongly type-II superconductor, we
may neglect the vector potential {{\bf A}} 
in the superfluid velocity ${\bf v}_s$. In
a singly quantized vortex $\Omega$ winds by $2\pi$ around the origin 
leading to a singularity
of the form ${\bf v}_s\simeq{1\over 2}\nabla\Omega =\hat\varphi/2r$.  
First, for the {\em holon} vortex we assume that
$\rho_b$ vanishes in the core as some power $\rho_b(r)\sim r^\nu$ and 
$\rho_\Delta(r)\approx\bar\rho_\Delta$ 
remains approximately constant. Eq.\ (\ref{gl:b}) then becomes
\begin{equation}
({1\over 4}-\nu^2)r^{\nu-2} + (r_b+v\bar\rho_\Delta^2)r^\nu + 
u_b\bar\rho_b^2r^{3\nu}=0,
\label{glcore}
\end{equation}
where we have neglected $\rho_b^2(r)$ compared to $4\bar\rho_\Delta^2$
in the denominator of the last term in Eq.\ (\ref{gl:b}). The most singular
term in Eq.\ (\ref{glcore}) is the first one and we must demand that the
coefficient of $r^{\nu-2}$ vanishes. This implies $\nu={1\over 2}$. The
asymptotic short distance behavior of the holon amplitude therefore
can be written as
\begin{equation}
\rho_b(r)\simeq c_b\bar\rho_b \left(r\over\xi_b\right)^{1/2},
\label{rhobcore}
\end{equation}
where $c_b$ is a constant of order unity which may be determined 
by the full integration of Eqs.\ (\ref{gl:all}).
Similar analysis of Eq.\ (\ref{gl:a}) in the vicinity of the {\em spinon} 
vortex yields 
\begin{equation}
\rho_\Delta(r)\simeq c_\Delta\bar\rho_\Delta \left(r\over\xi_d\right), 
\label{rhodcore}
\end{equation}
with $\rho_b$ approximately constant. 

We notice the different power laws
in the holon and spinon results. Operationally this difference arises from
different numerical prefactors of the respective superfluid velocity terms 
in Eqs.\ (\ref{gl:all}). Physically, the unusual $r$ dependence of the
holon amplitude in the core reflects the fact that the field $b$ describes
a condensate of single holons, each carrying charge $e$. Superconducting 
vortex with the flux quantum $\Phi_0$ represents a magnetic 
``half-flux'' for the 
holon field which results in non-analytic behavior of $\rho_b(r)$ at the 
origin. Singly quantized holon vortex is therefore a peculiar object and we
shall discuss it more fully in Section III. Here we note that the 
physical superconducting order parameter amplitude $|\Psi|=\rho_\Delta\rho_b^2$
remains analytic in the core of both the spinon and the holon
vortex.

\subsection{Holon vs. spinon vortex: the phase diagram}

We are now in the position to estimate the energies of the two types 
of vortices and deduce the phase diagram for the ``normal'' state in the 
vortex core. To this end we consider a single isolated vortex centered
at the origin.
The total vortex line energy can be divided into electromagnetic 
and core contributions\cite{fetter1}.
The electromagnetic contribution 
consists of the energy of the supercurrents and the magnetic 
field outside the core region. It may be estimated 
by assuming that the amplitudes $\rho_\Delta$ and $\rho_b$ 
have reached their bulk values $\bar\rho_\Delta$ and $\bar\rho_b$ respectively.
Taking curl of Eq.\ (\ref{lona}) and noting
that $\nabla\times\nabla\Omega=2\pi\delta({\bf r})$ for a singly quantized vortex
we obtain the London equation for the  magnetic field 
${\bf B}=\nabla\times{\bf A}$ of the form
\begin{equation}
B-\lambda^2\nabla^2B=\Phi_0\delta({\bf r})
\label{lon}
\end{equation}
where 
\begin{equation}
\lambda^{-2}=8\pi e^2 
{4\bar\rho_\Delta^2\bar\rho_b^2\over 4\bar\rho_\Delta^2+\bar\rho_b^2}.
\label{lam}
\end{equation}
has the meaning of the London penetration depth for the effective GL theory
(\ref{feff}). Aside from the unusual form of $\lambda$, Eq.\ (\ref{lon})
is identical to the conventional London equation.
The corresponding electromagnetic energy is therefore 
the same for both types of vortices
and can be calculated in the usual manner\cite{abrikosov1,fetter1,sachdev1} 
obtaining 
\begin{equation}
E_{\rm EM}\simeq\left({\Phi_0\over 4\pi\lambda}\right)^2 \ln\kappa,
\label{em}
\end{equation}
with $\kappa=\lambda/\max(\xi_\Delta,\xi_b)$ being the generalized GL ratio. 

To estimate the core contribution to the vortex line energy we assume that 
one of the amplitudes is suppressed to zero in the core
\begin{equation}
\rho_i(r)=0, \ \ \ r<\xi_i,
\label{rhor}
\end{equation}
while the other one stays constant and equal to its bulk value. This is a
very 
crude approximation which we justify below by an exact numerical computation.
With these assumptions, the core energy is 
\begin{equation}
E_{\rm core}^{(i)}\simeq\left({\Phi_0\over 4\pi\lambda_i}\right)^2, 
\label{ecore}
\end{equation}
where $i=\Delta,b$ for spinon and holon vortex respectively and 
\begin{equation}
\lambda^{-2}_i=8\pi e^2 \bar\rho_i^2.
\label{lami}
\end{equation}
Such a crude approximation overestimates the core energy. A more accurate
analysis\cite{abrikosov1,fetter1}, which we do not pursue here,
allows for a more realistic variation of $\rho_i(r)$ in the core and
indicates that the
value of $E_{\rm core}^{(i)}$ has the same form as Eq.\ (\ref{ecore})
multiplied by a numerical factor $c_1\approx 0.5$\cite{hu1,alama1}. 
Thus, the total energy
of the vortex line can be written as
\begin{equation}
E^{(i)}=\left({\Phi_0\over 4\pi\lambda}\right)^2\ln\kappa
+c_1\left({\Phi_0\over 4\pi\lambda_i}\right)^2, 
\label{evortex}
\end{equation}
where again $i=\Delta,b$ for spinon and holon vortex respectively.
Eq.\ (\ref{evortex}) parallels the Abrikosov expression for the vortex line 
energy in a conventional GL theory\cite{abrikosov1} 
where $\lambda$ and $\lambda_i$ are identical and equal to the ordinary 
London penetration depth.
\begin{figure}[t]
\epsfxsize=8.5cm
\epsffile{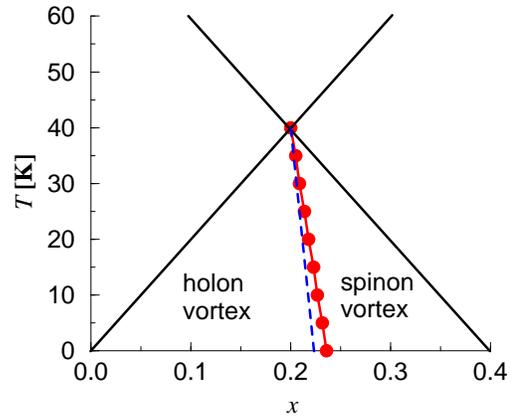}
\caption[]{Vortex core phase diagram for GL parameters chosen as follows:
$\alpha_\Delta=0.13$, $\alpha_b=0.10$, $T_0=200$K, $x_m=0.2$,
$u_\Delta=u_b=1.0$ and $v=0.5$.
Dashed line marks the phase boundary $T_g(x)$ obtained from 
Eq.\ (\ref{tg}) while the solid circles correspond to the 
numerical calculation with the same parameters. }
\label{fig2}
\end{figure}

In the vortex state described by the free energy (\ref{feff}) the vortex with
lower energy $E^{(i)}$ will be stabilized. Eq.\ (\ref{evortex}) implies that 
the difference in energy between the two types of vortices comes primarily
from the
core contribution, as expected on the basis of the physical argument presented
above. Condition $\lambda_\Delta=\lambda_b$ marks the 
transition point between the two solutions. For fixed GL parameters $T_0$, 
$x_m$, $\alpha_i$, $u_i$ and $v$ this defines a transition line in the $x$-$T$ 
plane. According to (\ref{lami}) the equation for this line is 
\begin{equation}
\bar\rho_\Delta(x,T)=\bar\rho_b(x,T).
\label{tran}
\end{equation}
Using Eqs.\ (\ref{ri}-\ref{rho}) one can obtain an explicit expression 
for the transition temperature $T_g$ between two types of vortices as
a function of doping
\begin{equation}
T_g(x)=T_0\left[{2x_m-x\over 1-\beta}+{x\over 1-\beta^{-1}}\right],
\label{tg}
\end{equation}
with 
\begin{equation}
\beta={\alpha_b(u_\Delta+v)\over \alpha_\Delta(u_b+v)}.
\label{beta}
\end{equation}
Eq.\ (\ref{tg}) describes a straight line in the $x$-$T$ plane, originating
at $[x_m,T_0x_m]$, i.e. maximal $T_c$ at optimum doping, and terminating at 
$[2x_m/(1+\beta),0]$. Generically, we expect that parameters $\alpha_i$ and
$u_i$ will be comparable in magnitude for the holon and spinon channels.
Parameter $\beta$ defined in Eq.\ (\ref{beta}) will therefore be of order
unity. The typical situation for $\beta=0.77$ 
is illustrated in Figure \ref{fig2}.
More generally the quartic coefficients $u_i$ and $v$ could exhibit weak
doping and temperature dependences leading to a curvature in the phase 
boundary. 

The appealing feature of the present theory is that parameter 
$\beta$ may vary from compound to compound. Thus, the experimental fact
that in BSCCO the pseudogap in the core persists
into the overdoped region is easily accounted for in the present theory. 
It would be interesting to see if the transition
from holon to spinon vortex as a function of doping could be experimentally 
observed. A good candidate for such observation would be LSCO, 
where the transport measurements in pulsed magnetic fields\cite{ando1}
established a metal-insulator transition around optimal doping, i.e.\
 $\beta\approx 1$. The current
theory predicts a holon vortex with the pseudogap spectrum in the
underdoped (insulating) region and spinon vortex with conventional metallic
spectrum on the overdoped side.

\subsection{Numerical results}

In order to put the above analytical estimates on firmer ground we now pursue
numerical computation of the vortex line energy. For simplicity we
consider the strongly type-II situation $(\kappa\gg1)$ where the vector
potential term in ${\bf v}_s$ can be neglected to an excellent approximation,
as long as we focus on the behavior close to the core.    
We are then faced with the task of numerically minimizing
the free energy (\ref{feff}) with respect to the two cylindrically
symmetric amplitudes $\rho_\Delta(r)$ and $\rho_b(r)$. As noted by Sachdev
\cite{sachdev1} direct numerical minimization of the free energy (\ref{feff})
provides a more robust solution than the numerical
integration of the coupled differential equations (\ref{gl:all}). 

We discretize the free energy functional (\ref{feff}) on a disk of a 
radius $R\gg\xi_i$ in the radial coordinate $r$ with up to $N= 2000$
spatial points. 
We then employ the Polak-Ribiere variant of the Conjugate Gradient Method
\cite{numrec} to minimize this discretized functional with respect to 
$\rho_\Delta(r_j)$ and $\rho_b(r_j)$, initialized to suitable single vortex trial 
functions. The procedure converges very rapidly and the results are 
insensitive to the detailed shape of the trial functions as long as they 
saturate to the correct bulk values outside the vortex core.

\begin{figure}[t]
\epsfxsize=8.5cm
\epsffile{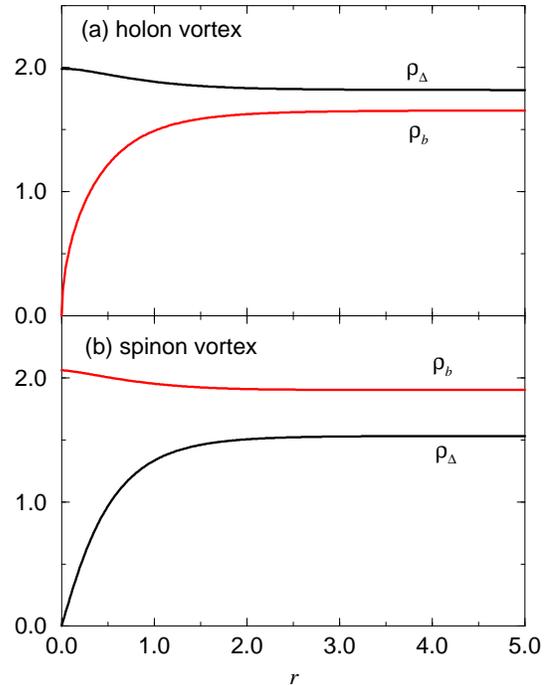}
\caption[]{Order parameter amplitudes near a single isolated vortex for
GL parameters specified in Figure \ref{fig2}.
The holon vortex is plotted for $T=0$ and
$x=0.22$ (implying coherence lengths $\xi_\Delta=0.63$ and $\xi_b=0.70$),
while the spinon vortex is plotted for $T=0$ and $x=0.24$ (implying
$\xi_\Delta=0.75$ and $\xi_b=0.60$).}
\label{fig3}
\end{figure}

Typical results of our numerical computations are displayed in Figure
(\ref{fig3}) and are in complete agreement with the analytical
considerations of the preceding subsections. Note in particular that 
$\rho_b(r)$ in the holon vortex  vanishes with infinite slope, consistent
with Eq.\ (\ref{rhobcore}). Plotting $\rho_b^2(r)$ confirms that the 
exponent is indeed $1/2$. In the spinon vortex
$\rho_\Delta(r)$ is seen to vanish linearly as expected on the basis of
Eq.\ (\ref{rhodcore}). The nonvanishing order parameter is slightly 
elevated in the core reflecting the effective ``repulsion'' between the
two amplitudes contained in the $v$-term of the free energy. The results
for the spinon vortex are consistent with those of Ref.\ \cite{sachdev1}.

We explored a number of other parameter configurations and obtained 
similar results.
We find that Eq.\ (\ref{tran}) is a good predictor of the 
transition line between the holon and spinon vortex, although the precise
numerical value of the transition temperature $T_g$ for given $x$ tends
to deviate slightly from the value predicted by Eq.\ (\ref{tg}). This is
illustrated in Figure (\ref{fig2}) where we compare the vortex core
phase diagrams obtained numerically and from Eq.\ (\ref{tg}). Interestingly,
the deviation always tends to enlarge the holon vortex sector of the phase
diagram at the expense of the spinon vortex sector. This is presumably 
because the sharper $\sim\sqrt{r}$ suppression of the holon order parameter
in the core costs less condensation energy.

\section{Gauge fluctuations and the spectral properties in the core}

Theory of the vortex core based on the effective action (\ref{feff})
appears to yield results consistent with the STM data on cuprates
\cite{renner1,pan2} in that it implies stable holon vortex solution over
the large portion of the superconducting phase diagram. The state inside the 
core of such a holon vortex is characterized by vanishing amplitude of the
holon condensate field,
$|b|=0$, and a finite spin gap $|\Delta|\approx\Delta_{\rm bulk}$. This is the 
same state as in the pseudogap region above $T_c$. One would thus expect
the electronic spectrum in the core to be similar to that found in the normal 
state of the underdoped cuprates, in agreement with the 
data\cite{renner1,pan2}. The holon vortex
with this property carries conventional superconducting flux quantum 
$\Phi_0$, in accord with experiment. This general agreement between theory
and experiment would suggest that the effective action (\ref{feff}) 
provides the sought for phenomenological description of the vortex core 
physics in cuprates.
In what follows we amplify our argumentation that it is also
tenable in a broader theoretical context in that it naturally
follows from the U(1) slave boson models extensively studied in the 
classic and more recent high-$T_c$ literature. We then provide a more detailed
discussion of the vortex core spectra and propose an explanation for the
experimentally observed core bound states.

\subsection{Significance of the $f_{\rm gauge}$ term} 

Derivation of the effective action (\ref{feff}) from the more general U(1)
action (\ref{fgl1}) hinges on our assumption that the stiffness $\sigma$ 
of the 
gauge field ${\bf a}$ is low and that the $f_{\rm gauge}$ term (\ref{fgauge})
can be neglected. Assumption of large $\sigma$ by SNL leads to very different
vortex solutions\cite{sachdev1,lee3} which appear inconsistent with the recent
experimental data. We first expand on our discussion as to why is  
$f_{\rm gauge}$ term important and then we argue why it may be permissible
to neglect it in the realistic models of cuprates. 

To facilitate the discussion let us rewrite Eq.\ (\ref{fgl1}) by resolving
the complex matter fields into amplitude and phase components:
\begin{eqnarray}
f_{\rm GL}&=& f_{\rm amp} +\rho_\Delta^2(\nabla\phi -2{\bf a})^2
+\rho_b^2(\nabla\theta-{\bf a}-e{\bf A})^2 \nonumber \\
&+&{1\over 8\pi}(\nabla\times{\bf A})^2 + 
{\sigma\over2}(\nabla\times{\bf a})^2,
\label{fgl2}
\end{eqnarray}
with $f_{\rm amp}$ specified by Eq.\ (\ref{famp}). Now consider situation
in which the sample is subjected to uniform magnetic field 
${\bf B}=\nabla\times{\bf A}$. Two scenarios (discussed previously by SNL) 
appear possible. In the first, the internal gauge field develops no net 
flux, $\langle\nabla\times{\bf a}\rangle=0$, and the holon phase 
$\theta$ develops 
singularities in response to ${\bf A}$ such that 
$$\nabla\times\nabla\theta=2\pi\sum_j\delta({\bf r}-{\bf r}_j),$$ 
where ${\bf r}_j$ denotes the vortex positions. The holon amplitude $\rho_b$
is driven to zero at ${\bf r}_j$, essentially to prevent the free energy from
diverging due to the singularity in the phase gradient.
Since holons carry charge $e$, each vortex is threaded by 
flux $hc/e$, i.e.\ twice the superconducting flux quantum $\Phi_0=hc/2e$. 
This solution represents the doubly quantized holon vortex lattice, considered
by SNL. 

In the second scenario
${\bf a}$ develops a net flux such that ${\bf a}\approx -e{\bf A}$, which screens out the
${\bf A}$ field in the holon term but produces a net flux $-2e{\bf A}$ in the 
spinon term. In response to this flux, spinon phase $\phi$ develops 
singularities such that 
$$\nabla\times\nabla\phi=2\pi\sum_j\delta
({\bf r}-\tilde{\bf r}_j),$$
corresponding to the spinon vortex lattice. $\tilde{\bf r}_j$ denotes vortex
positions which will be different from ${\bf r}_j$ since at the fixed field
$B$ there will be twice as many spinon vortices as holon vortices. 
(Spinon vortices carry conventional superconducting quantum of flux $\Phi_0$.) 
In this case $\rho_\Delta$ is driven to zero at the vortex centers.
In this scenario 
one pays a penalty for nucleating the net flux in $\nabla\times{\bf a}$ due
to last term in Eq.\ (\ref{fgl2}). This energy cost can be estimated as
\begin{equation}
E_\sigma\simeq 8\pi\sigma e^2 \left({\Phi_0\over 4\pi\lambda}\right)^2
\label{esig}
\end{equation}
per vortex. Stiffness $\sigma$ must be small enough so that $E_\sigma$ is
small compared to the vortex energy (\ref{evortex}). Taking the dominant 
$E_{\rm EM}$ term and neglecting $\ln\kappa$ this implies that
\begin{equation}
\sigma\ll {1\over 8\pi e^2},
\label{sig}
\end{equation}
which is the same condition as considered in Ref.\ \cite{sachdev1}.

Now consider a {\em third} scenario in which a {\em singly 
quantized} holon vortex emerges. As a starting point consider the spinon 
vortex solution just described. In the underdoped regime the amplitude piece
$f_{\rm amp}$ would favor suppressing the holon amplitude in the core 
instead of the spinon amplitude but according to our previous considerations
this would ordinarily require formation of a doubly quantized vortex whose 
magnetic energy is too large. However, if the gauge field stiffness $\sigma$
is sufficiently small, the system could lower its free energy by
setting up singularities in ${\bf a}$ which would precisely cancel the 
singularities in $\nabla\phi$ and shift them to the holon term. 
To arrive at this situation imagine
contracting the initially uniform flux $\nabla\times{\bf a}$ so that it
becomes localized in the individual vortex core regions. Taking this 
procedure to the extreme, i.e. taking the limit $\sigma\to 0$,
the gauge field will form ``flux spikes'' of the form 
\begin{equation}
2(\nabla\times{\bf a})
=-\nabla\times\nabla\phi=-2\pi\sum_j\delta({\bf r}-\tilde{\bf r}_j),
\label{sing}
\end{equation}
completely localized at the vortex centers. 
Gauge field of this form indeed completely cancels the singularities in the 
spinon
phase gradient in Eq.\ (\ref{fgl2}) and $\rho_\Delta$ is no longer forced
to vanish in the core. The singularities now appear in the holon term,
but they stem from ${\bf a}$ rather that $\nabla\theta$ which remains 
nonsingular. Consequently, $\rho_b$ is forced to vanish in the vortex cores. 
By construction the vortices are located at $\tilde{\bf r}_j$  and 
are therefore singly quantized.
This is the singly quantized holon vortex 
discussed in the framework of the free energy (\ref{feff}). Based on the
above discussion the singly quantized holon vortex can be thought of as a
composite object formed by attaching half quantum $(h/2)$ of the  
fictitious gauge flux $\nabla\times{\bf a}$ to the spinon vortex. Within
the full compact U(1) theory this is essentially equivalent to the Z$_2$
vortex discussed by Wen\cite{wen2} in the framework of topological orders
in spin liquids. 

In the framework of the free energy (\ref{fgl2}) one pays a penalty for such 
a singular solution due to the gauge stiffness term. In the present 
continuum model
this penalty per single vortex is actually infinite, since according to 
Eq.\ (\ref{sing}) it involves a 
spatial integral over $[\delta({\bf r}-\tilde{\bf r}_j)]^2$. Thus, in the
continuum model
the singular solutions of this type are prohibited. In reality, however,
we have to recall that our effective action (\ref{fgl1}) descended from
a microscopic 
lattice model for spinons and holons in which the gauge field ${\bf a}$ lives
on the nearest neighbor bonds of the ionic lattice. The ionic lattice constant
$d$ therefore provides a natural short distance cutoff and the delta
function in Eq.\ (\ref{sing}) should be interpreted as a flux quantum $\Phi_0$
piercing an elementary plaquette of the lattice. The energy cost per vortex
thus becomes finite and is given by 
\begin{equation}
E'_\sigma\simeq {\sigma e^2\over 2} \left({\Phi_0\over d}\right)^2.
\label{esigp}
\end{equation}
Again, for the solution to be stable, $E'_\sigma$ must be negligible compared
to the vortex energy (\ref{evortex}). This implies 
\begin{equation}
\sigma\ll {1\over 8\pi^2e^2} \left({d\over \lambda}\right)^2,
\label{sigp}
\end{equation}
which is a much more stringent condition than (\ref{sig}) since in cuprates
$d\ll\lambda$. 

When condition (\ref{sigp}) is satisfied it is permissible to neglect 
the $f_{\rm gauge}$ term in the effective action (\ref{fgl1}) and
it becomes fully equivalent to (\ref{feff}) as far as the vortex solutions are 
concerned. 
Eq.\ (\ref{sigp}) gives the precise meaning to the requirement of the weak 
stiffness of the gauge field loosely stated when deriving the effective
action (\ref{feff}).

\subsection{Microscopic considerations}

As mentioned in the introduction, the gauge field ${\bf a}$ has no dynamics
in the original U(1) microscopic model, as it only serves
to enforce a constraint on spinons and holons. The stiffness term 
(\ref{fgauge}) in the effective
theory was assumed to arise in the process of integrating out the
microscopic degrees of freedom\cite{sachdev1,lee3}. While such term is
certainly permitted by symmetry, assessing its strength
$\sigma$ is a nontrivial issue since even deep in the superconducting phase
neither holons nor spinons are truly gapped. Thus, in general, integrating
out these degrees of freedom may lead to singular and nonlocal interactions
between the condensate and the gauge fields. To our knowledge the procedure
has not been explicitly performed for the U(1) model and the precise form or
magnitude of the gauge stiffness term is unknown. General 
considerations\cite{nayak1} suggest
that the gauge stiffness term is negligible in the class 
of models with exact local U(1) symmetry connecting the phases of holons 
and spinons.

Consider now an intermediate representation of the problem where only
high energy microscopic degrees of freedom have been integrated out. In the 
presence of a cutoff this is a well defined procedure even for gapless
excitations, as explicitly shown by Kwon and Dorsey\cite{kwon1} for a 
simple BCS model. The corresponding effective Lagrangian density of the 
present U(1) model can be written as
\begin{eqnarray}
{\cal L}_{\rm eff}&=&
{\kappa_\Delta^\mu\over 2}(\partial_\mu\phi-2a_\mu)^2
+{\kappa_b^\mu\over 2}(\partial_\mu\theta-a_\mu-eA_\mu)^2
-f_{\rm amp} \nonumber \\
&+&(\partial_\mu\phi-2a_\mu)J_{\rm sp}^\mu
+(\partial_\mu\theta-a_\mu-eA_\mu)J_h^\mu \nonumber \\ 
&+&{\cal L}_{\rm sp}[\psi_{\rm sp},\psi_{\rm sp}^\dagger;\rho_\Delta]
+{\cal L}_h[\psi_h,\psi_h^\dagger;\rho_b] +{\cal L}_{\rm EM}[A_\mu].
\label{leff}
\end{eqnarray}
The Greek index $\mu$ runs over time and two spatial dimensions, 
$\kappa_i^0$ are compressibilities of the holon and spinon condensates,
while 
\begin{equation}
\kappa_i^j=-2(\rho_i)^2,\ \ \  i=\Delta, b, \ \ \  j=1,2, 
\label{kappa}
\end{equation}
are the
respective phase stiffnesses. $J_{\rm sp}^\mu$ and $J_h^\mu$ are spinon and
holon three currents respectively and ${\cal L}_{\rm sp}$ and 
${\cal L}_h$ are the low energy effective Lagrangians for 
the fermionic spinon field $\psi_{\rm sp}$ and bosonic holon field $\psi_h$. 
${\cal L}_{\rm EM}$ is the Maxwell Lagrangian for the physical electromagnetic
field. Thus, ${\cal L}_{\rm eff}$ describes
an effective low energy theory of spinons and holons coupled to their 
respective collective modes and a fluctuating U(1) gauge field. Similar theory 
has been recently considered by Lee\cite{dhlee1}. 
 
The precise form of the microscopic Lagrangians  ${\cal L}_{\rm sp}$ and 
${\cal L}_h$ is not important for our discussion. The salient feature which 
we exploit here is that only the amplitude of the respective condensate field 
enters into ${\cal L}_{\rm sp}$ and ${\cal L}_h$. Coupling to the phases and
the gauge field is contained entirely in the Doppler shift terms [second line
of Eq.\ (\ref{leff})]. Such form of the coupling is largely dictated by the 
requirements of the gauge invariance and the particular form Eq.\ (\ref{leff})
can be explicitly derived by gauging away the respective phase factors
from the $\psi$ fields\cite{balents1,kwon1}. 

The gauge field $a_\mu$ enters the effective Lagrangian (\ref{leff}) only via
two gauge invariant terms: $(\partial_\mu\phi-2a_\mu)$ and 
$(\partial_\mu\theta-a_\mu-eA_\mu)$, which may be interpreted as the three 
velocities of the spinon and holon condensates respectively. 
Furthermore, the only coupling between
holons and spinons arises from $a_\mu$. Therefore, if we now proceed to 
integrate out the remaining microscopic degrees of freedom from 
${\cal L}_{\rm eff}$, the two 
velocity terms will not mix. This consideration suggests that upon 
integrating out all of the microscopic degrees of freedom, the resulting
gauge stiffness term will be of the form
\begin{eqnarray}
f'_{\rm gauge}&=&{\sigma_\Delta\over2}[\nabla\times(2{\bf a}-\nabla\phi)]^2
\nonumber \\
&+&{\sigma_b\over2}[\nabla\times({\bf a}+e{\bf A}-\nabla\theta)]^2.
\label{fgaugep}
\end{eqnarray}
Clearly, such term is permitted by the gauge symmetry. Furthermore, we note 
that for smooth (i.e.\ vortex free) configurations of phases the gradient
terms will contribute nothing and we recover the gauge term considered
in Ref.\ \cite{lee3}. 

In the presence of a vortex in $\phi$ or $\theta$ the $f'_{\rm gauge}$
term will contribute formally divergent energy. Regularizing
this on the lattice, as discussed above Eq.\ (\ref{esigp}),
this energy will become finite and can be interpreted simply as the
energy of the spinon or holon vortex core states, which have been integrated 
out. In the microscopic theory (\ref{leff}) such energy would arise upon
solving the relevant fermionic or bosonic vortex problem. 

We stress that, as concluded
in the preceding subsection, the main theoretical obstacle to the 
formation of a singly quantized holon vortex in the original SNL theory 
was the appearance of a formally
divergent contribution in the $f_{\rm gauge}$ term (\ref{fgauge}). The 
argument above suggests that $f_{\rm gauge}$ in Eq.\ (\ref{fgl1})
 should be replaced by Eq.\ 
(\ref{fgaugep}), in which  such formally divergent contribution appears for
{\em arbitrary} vortex configuration and upon regularization has a simple
physical interpretation in terms of the energy of the vortex core states. 
Usage of the physically
motivated term (\ref{fgaugep}) in place of (\ref{fgauge}) therefore 
removes the bias against the singly quantized holon vortex solution, which 
appears to be realized in real materials. With (\ref{fgaugep}) any bias 
between the holon and spinon vortex solutions can result only from the 
difference between the two stiffness constants $\sigma_\Delta$ and 
$\sigma_b$. It is reasonable on physical grounds to assume that
constants $\sigma_\Delta$ and $\sigma_b$ are of the similar magnitudes. 
Furthermore, on the basis of Ref.\ \cite{nayak1} we expect these constants
to be negligibly small in the physically relevant models. 
Consequently we
expect that neglecting the $f_{\rm gauge}$ term as in our derivation
of effective action (\ref{feff}) will result in accurate determination
of the phase diagram for the state in the vortex core.

\subsection{Vortex core states}

The phenomenological theory based on the effective action (\ref{feff}) 
does not allow us to address the interesting question of the nature of 
the fermionic states in the vortex core. To do this we need to consider 
the microscopic  Lagrangian density
(\ref{leff}). While the fully self consistent calculation is likely 
to be prohibitively difficult, one can obtain qualitative insights by first
solving the GL theory (\ref{feff}) as described in Sec.\ II, and then
using the order parameters $\rho_\Delta$ and $\rho_b$ 
as an input to the fermionic and bosonic sectors 
of the theory specified by Eq.\ (\ref{leff}). The work on a detailed
solution of this type is in progress. Here we wish to point out some 
interesting features of such a theory and argue that it may indeed exhibit
structure in the low energy spectral density similar to that found 
experimentally\cite{maggio1,pan2}.

It is instructive to integrate out the gauge fluctuations from the 
Lagrangian (\ref{leff}) as first discussed by Lee\cite{dhlee1}. Since
${\cal L}_{\rm eff}$ is quadratic in $a_\mu$ the integration can be
explicitly performed resulting in the Lagrangian of the form
\begin{eqnarray}
{\cal L}'_{\rm eff}&=&
{1\over 2}K_\mu(v_s^\mu)^2 -f_{\rm amp} + {\cal L}_{\rm EM}\nonumber \\
&-& {2\kappa_b^\mu\over 4\kappa_\Delta^\mu+\kappa_b^\mu}
(v_s^\mu J_{\rm sp}^\mu)
+{4\kappa_\Delta^\mu\over 4\kappa_\Delta^\mu+\kappa_b^\mu}
(v_s^\mu J_h^\mu)\nonumber \\
&+&{\cal L}_{\rm sp} +{\cal L}_h
-{1\over 2}{1\over 4\kappa_\Delta^\mu+\kappa_b^\mu}(2J_{\rm sp}^\mu+J_h^\mu)^2,
\label{leffi}
\end{eqnarray}
where $K_\mu=4\kappa_\Delta^\mu\kappa_b^\mu/(4\kappa_\Delta^\mu+\kappa_b^\mu)$
and 
\begin{equation}
v_s^\mu=(\partial_\mu\theta-{1\over2}\partial_\mu\phi-eA_\mu)
\label{vs}
\end{equation}
is
the physical superfluid velocity. The first line reproduces the GL effective
action (\ref{feff}) for the condensate fields, 
the second line describes the Doppler shift coupling 
of the superfluid velocity to the microscopic currents, and the  third line
contains spinon and holon pieces with additional current-current interactions
generated by the gauge fluctuations\cite{dhlee1}.

We now discuss the physical implications of Eq.\ (\ref{leffi}) for the two 
types of vortices. We focus
on the static solutions (i.e. we ignore the time dependences of various 
quantities, e.g.\ taking $v_s^0=0$) of ${\cal L}'_{\rm eff}$ in the 
presence of a single isolated vortex. We are interested in the local
spectral function of a physical electron. This is
given by a convolution in the energy variable of the spinon and holon 
spectral functions. According
to the analysis presented in Ref.\ \cite{lee2}, at low temperatures 
the electron spectral function will be essentially equal to the spinon
spectral function. Convolution with the
holon spectral function which is dominated by the sharp coherent peak due to 
the condensate merely leads
to a small broadening of the order $T$. In the following we therefore focus
on the behavior of spinons in the vicinity of the two types of vortices.

By inspecting Eq.\ (\ref{leffi}) it is easy to see that the excitations
inside the {\em spinon vortex} will be qualitatively 
very similar to those found in the conventional vortex described by the weak
coupling $d$-wave BCS theory\cite{wang1,franz1,kita1}. In particular according 
to Eq. (\ref{rhodcore}) we have $\kappa_\Delta\sim r^2$,
and $\kappa_b\sim$ const in the core. Recalling furthermore that 
$|{\bf v}_s|\sim 1/r$ we observe that the spinon current ${\bf J}_{\rm sp}$ 
is coupled to a term that diverges as $1/r$ in the core (just as in a 
conventional vortex), while the holon current ${\bf J}_h$ is coupled to a 
nonsingular term. Thus, one may conclude that holons remain essentially
unperturbed by the phase singularity in the spinon vortex while the spinons 
obey the essentially conventional Bogoliubov-de Gennes equations for a 
$d$-wave vortex. 

In the {\em holon vortex} the situation is quite different. According to  
Eq. (\ref{rhobcore}) we have $\kappa_b\sim r$
and $\kappa_\Delta\sim$ const in the core. The spinon current 
${\bf J}_{\rm sp}$ is now
coupled to a nonsingular term ($1/r$ divergence in $v_s$ is canceled by 
$\kappa_b\sim r$).
Therefore, there will be no topological perturbation in the spinon sector
and we expect the spinon wavefunctions to be essentially unperturbed by 
the diverging superfluid velocity. Spinon spectral density in the core 
should be qualitatively similar to that far outside the core. This is our basis
for expecting a pseudogap-like spectrum in the core of a holon vortex. 

We now address the possible origin of the experimentally observed vortex
core states\cite{maggio1,pan2} within the present scenario for a holon vortex.
To this end consider the effect of the last term in Eq.\ (\ref{leffi}) which 
we ignored so far. Upon expanding the binomial 
the temporal component is seen to contain a density-density 
interaction of the form $J_{\rm sp}^0 J_h^0$ where $J_h^0$ is the local
density of {\em uncondensed} holons. Since the holon order parameter
vanishes in the core and the electric neutrality dictates that 
the total density of holons must be approximately constant in space, we 
expect that uncondesed holon density will behave roughly as
\[
J_h^0(r)=\bar\rho_b-\rho_b(r);
\]
$J_h^0(r)$ will have a spike in the core of a holon vortex.
Insofar as $J_h^0(r)$ can be viewed as a static potential acting on
spinons, the uncondensed holons in the vortex core can be thought of as 
creating a scattering potential, akin to an 
impurity embedded in a  $d$-wave superconductor. 
In fact, formally the spinon problem is identical to the problem of a 
fermionic quasiparticle in a $d$-wave superconductor
in zero field in the presence of a localized impurity potential. It is known
that such problem exhibits a pair of marginally bound impurity 
states\cite{balatsky1}
at low energies which result in sharp resonances in the spectral density
inside the gap. Such states have been extensively studied theoretically
\cite{balatsky2,flatte1,shnirman1,atkinson1} and their existence was 
recently confirmed  experimentally by Pan {\em et al.}
\cite{pan1}. We propose here that, within the formalism of Eq.\ (\ref{leffi}),
the same mechanism could give rise to the low energy quasiparticle states 
in the core of a holon vortex. Such structure, if indeed confirmed by a
microscopic calculation, could explain the spectral features observed
experimentally in the vortex cores of cuprate superconductors
\cite{maggio1,pan2}.

\section{Conclusions}

Scanning tunneling spectroscopy of the vortex cores affords a unique 
opportunity for probing the underlying ``normal'' ground state in cuprate 
superconductors. The existing experimental data
on YBCO and BSCCO strongly suggest that conventional mean field weak coupling
theories \cite{soininen1,wang1,franz1,kita1,volovik1,maki1,ichioka1}
fail to describe the physics of the vortex core. 
Our main objective was to develop a theoretical framework for understanding 
these spectra and the nature of the strongly
correlated electronic system which emerges once the superconducting 
order is suppressed. We have shown that phenomenological model (\ref{fgl1})
based on a variant of the U(1) gauge field slave boson theory
\cite{dhlee1} contains the right physics, provided that the gauge field 
stiffness
is vanishingly small. The latter assumption is consistent with the 
general arguments involving local gauge symmetry\cite{nayak1}. In such a 
theory the gauge field can be explicitly integrated out, resulting in the 
effective action (\ref{feff}) which contains one phase degree of freedom 
representing the phase of a Cooper pair and two amplitude degrees of
freedom representing the holon and spinon condensates. 

Analysis of the effective theory (\ref{feff}) in the presence of a 
magnetic field
establishes existence of two types of vortices, spinon and holon, with 
contrasting spectral properties in their core regions.
Our holon vortex is singly quantized and therefore differs in a profound way
from the doubly quantized holon vortex discussed by SNL\cite{sachdev1,lee3}.
As indicated in Figure \ref{fig2} such a singly quantized 
holon vortex is expected to be 
stable over the large portion of the phase diagram on the underdoped side. 
Quasiparticle spectrum in the core of a holon vortex is predicted to exhibit 
a ``pseudogap'', similar to that found in the underdoped normal region 
above $T_c$. This is consistent with the data of Renner {\em et al.}
\cite{renner1} who pointed out a remarkable similarity between the vortex
core and the normal state spectra in BSCCO. Spinon vortex, on the other hand,  
should be virtually indistinguishable from the conventional $d$-wave BCS
vortex and is expected to occur on the overdoped side of the phase diagram. 
Transition from the insulating holon vortex to the metallic spinon vortex
as a function of doping is a concrete testable prediction of the present 
theory. 

Phenomenological theory based on the effective 
action (\ref{feff}) does not permit 
explicit evaluation of the electronic spectral function. To this end
we have considered the corresponding microscopic theory 
(\ref{leffi}) and concluded that
holon vortex will indeed exhibit a pseudogap like spectrum. Such qualitative
analysis furthermore suggests a plausible mechanism for the sharp vortex 
core states observed in YBCO\cite{maggio1}  and BSCCO
\cite{pan2}. We stress that conventional mean field weak coupling theories 
yield neither pseudogap nor the core states.
In the core of a holon vortex such states will arise
as a result of spinons scattering off of the locally uncondensed holons,
in a manner analogous to the quasiparticle resonant states in the vicinity
of an impurity in a $d$-wave superconductor
\cite{balatsky1,balatsky2,flatte1,shnirman1,atkinson1}.
The latter conclusion is somewhat speculative and must be 
confirmed by explicitly solving the fermionic sector of the microscopic
theory (\ref{leffi}). 

On a broader theoretical front the importance of the vortex core spectroscopy
as a window to the normal state in the $T\to 0$ limit lies in its potential
to discriminate between various microscopic theories of cuprates. 
It is reasonable to assume that the observed pseudogap in the 
vortex core reflects the same physics as the pseudogap observed 
in the normal state.
This means that the mechanism responsible for the pseudogap must be operative 
on extremely short lengthscales, of order of several lattice spacings.
The U(1) slave boson theory considered in this work apparently satisfies 
this requirement. Obtaining the correct vortex core spectral functions could
serve as an interesting test for other theoretical approaches describing 
the physics of the underdoped cuprates\cite{lee2,pines1,levin1}.

It will be of interest to explore 
the implications of the effective theories (\ref{feff}) and (\ref{leffi}) 
in other physical situations. Of special interest are situations where 
the holon condensate amplitude is suppressed, locally
or globally, giving rise to ``normal'' transport properties (vanishing 
superfluid density) but quasiparticle excitations that are characteristic 
of a superconducting state. These include
the spectra in the vicinity of an impurity, 
twin boundary or a sample edge. In the latter case one might hope to observe 
a signature of the zero bias tunneling peak anomaly
(normally seen for certain geometries deep in the superconducting phase in 
the optimally doped cuprates) even above $T_c$ in the underdoped samples. 

\acknowledgments
The authors are indebted to A. J. Berlinsky, 
J. C. Davis, \O. Fischer, C. Kallin, D.-H. Lee, P. A. Lee,
S.-H. Pan, C. Renner, J. Ye and S.C. Zhang
for helpful discussions. This research was 
supported in part by NSF grant DMR-9415549 and by Aspen Center for Physics
where part of the work was done. 

\vskip 10pt
{\em Note added in proof.} After submission of this manuscript we learned
about complementary microscopic treatments of the spin-charge separated
state in the 
vortex core within U(1) \cite{han1} and SU(2) \cite{wen1} slave boson 
theories. The former agrees qualitatively with our phenomenological theory.  
Ref.\ \cite{wen1} proposes a new type of vortex which takes advantage of
the larger symmetry group SU(2). In a related development Senthil
and Fisher \cite{senthil1} discussed a Z$_2$ vortex (which is essentially 
equivalent to our singly quantized holon vortex) and proposed a 
``vison detection'' experiment based on trapping such a vortex in the hole 
fabricated in a strongly 
underdoped superconductor. Here we wish to point out that the experiment
will produce the same general outcome in a system described by the U(1) 
theory where the role of a vison will be played by a flux quantum of the
fictitious gauge field ${\bf a}$.

\end{document}